\documentclass[aps,12pt,showpacs]{revtex4}
\usepackage{graphicx}
\usepackage[english]{babel}
\usepackage {bm}

\def\h{\hbar}
\def\s{\sigma}

\begin{document}
\title{Exact two particle spectrum of the Heisenberg-Peierls chain}
\author{Julio Abad and J. G. Esteve }
\affiliation{
 Departamento de F\'{\i}sica Te\'orica,
 Facultad de Ciencias,
 Universidad de
 Zaragoza, 50009 Zaragoza, Spain.} 
\begin{abstract}
 The exact solution for the
 two particle spectrum
 of the Heisenberg-Peierls
 one dimensional spin chain is given by 
working in the fermionic representation. The resulting equations
 for the eigenvalues are,
 in some sense, similar to those of the Richardson's solution of the 
 BCS model and must be solved numerically.

\end{abstract}
\maketitle
%
%
%

\section{Introduction}

Low dimensional spin systems have attracted a great deal of interest
 over the last years because, both, theoretical and experimental reasons.
 In fact, from the experimental point of view, there are actually
 many quasi one dimensional compounds such as the organic series
 $(BCPTTF)_2X$ with $(X=AsF_6,\, PF_6)$, the cuprate $CuGeO_3$ or the
 $(VO)_2P_2O_7$ compounds, as well as other series as the 
 $TTFMS_4C_4(CF_3)_4$ with $M=Cu, Au, Pt,$ or $Ni$, which can be 
 well described as spin-Peierls chains \cite{spchain}. Furthermore, we have 
 now the evidence that the two dimensional antiferromagnetism is one of the
 crucial components of the high temperature
 superconductivity \cite{htc}, and we have also Heisenberg 
 spin-ladder compounds which can be viewed as intermediaries between one and
 two dimensions, and also as between half-integer $(S=1/2)$ and integer
 $(S=0,\, 1)$ spin chains.
 
From the theoretical side, these systems are interesting
 since
 their low dimensionality makes that  their properties are  strongly
 affected by quantum fluctuations and also because of 
 its apparent simplicity.
 For example, low dimensional electronic
 materials are known to be very sensitive
 to structural distortions driven by the electron-phonon interactions which
 breaks the symmetry of the original (i.e. without electron-phonon 
 interaction)  ground state, resulting in a less symmetric 
 but lower energy state, in which the electrons and ions are shifted, from
 their symmetric positions, in a regular manner that creates a periodic
 variation of the charge density called charge density wave.
 This is the famous Peierls instability which 
 opens a gap at the Fermi surface of the  one-dimensional electronic
 chain, transforming  what would be a metal (in absence of electron-phonon
 interactions) in a semiconductor. A similar effect is expected for
  one dimensional spin chains which are unstable against
 a dimerized spin-singlet ground state .

  One of  the most widely used  model for spin interactions is
 the generalized one dimensional
 Heisenberg Hamiltonian which 
 can be written, in terms of the quantum spin operators ${\bf S}_i = 
 (1/2)\, {\bm \sigma}_i$ (where ${\bm \sigma}_i$ are the Pauli matrices 
at the site $i$), as
\begin{eqnarray}
H= 4 \sum_{i=1}^M J_{i,i+1} {\bf S}_i {\bf S}_{i+1},
\label{heis}
\end{eqnarray}
and that, after some transformations, can be seen also as a one
 dimensional Hubbard model for electronic interactions.
The case when the coupling $J_{i,i+1}$ between
 nearest spins is taken as a constant
 is exactly soluble after Bethe and his celebrated Bethe ansatz \cite{r3}.
Later it was solved again   by applying 
 the Quantum Inverse Scattering methods, with which this model 
results inside of a family of solvable models \cite{r4}. 
However the Heisenberg model must be seen only as a first approximation 
to the real interactions, as long as a constant coupling between
 nearest-neighbors implies that the atoms are taken at fixed
 positions.
 If we account for the vibrations of the atoms,
 then the coupling between two nearest spins is proportional
 to the exchange integral and then it  can be written, up to first order, as
 $J_{n,n+1}=u -J_1 (v_n-v_{n+1})$ where $u$ and $J_1$ are constants
 and $v_n$ is the displacement of the nth atom from
 its equilibrium position. In the Born-Oppenheimer approximation, 
 $v_n = (-1)^n v_0$ (adiabatic limit) 
the Hamiltonian (\ref{heis}) can be written as
\begin{eqnarray}
H=\sum_{n=1}^{M}\left( u+ (-1)^n v \right) {\bm \sigma_n}
{\bm \sigma_{n+1}},
\label{hp}
\end{eqnarray}
with $v=2 J_1 v_0$. The  Pauli matrices ${\bm \s}_n$ act
 on the two-dimensional
 space $\eta_n={\cal C}^2$ whereas
$H$ acts on the $2^M$-dimensional Hilbert space,
\begin{eqnarray}
{\cal H}=\otimes_{n=1}^{M} \eta_n,
\label{iii}
\end{eqnarray}
which is the tensor product of $M$ two-dimensional ${\cal C}^2$ spaces.
We use  periodic boundary conditions  on the lattice with
$M=$ even playing the role of the length of the chain. The total
third component of the spin,
\begin{eqnarray}
S_3 = \sum_{n=1}^{M} \s^3_n
\label{iv}
\end{eqnarray}
is a conserved magnitude that we shall use to describe the
 states of the system. In general, 
to find the exact spectrum of the Hamiltonian (\ref{hp})
 is a difficult task and several methods like bosonization
 \cite{r5} and variational \cite{r6}  have been applied and
 actually for $v \ll |u|$, case of weak
dimerization, the low energy region is believed to be
 equivalent to that of the Gaussian model
\cite{gog}.
In this paper we want to find the low spin excitations
 over the state where all $M$ spins are aligned and
 pointing in the $-Z$ direction (note that the particular 
direction in which the spins are aligned is not relevant due to the
 rotational invariance   of the Hamiltonian). To do this we shall
 fermionize the problem with the Jordan-Wigner transformation
 and use a method analogous to that of the Richardson's solution
 \cite{rich}
 of the BCS model of superconductivity.

\section{one particle states}

As said above, our starting point will be an eigenstate of $H$
 with all spins aligned in the $-Z$ direction so that it is also an 
 eigenstate of $S_3$ with eigenvalue $-M$; starting from it, we want
 to find the 
 eigenstates of the Hamiltonian that have one spin flipped in the
 $+Z$ direction and  we call them the one-particle states.
 The first task is
 to transform the Hamiltonian (\ref{hp}) in a fermionic Hamiltonian
 by means of the Jordan-Wigner transformation since, although this is 
 not strictly necessary for these one-particle  states, it will be need
 for the 
 two-particle states, so we start by defining the fermionic operators:
\begin{eqnarray}
a_l&=&K(l) \s^{-}_l\,,\\
a^{\dagger}_l&=& K(l)\s^{+}_l,
\label{v}
\end{eqnarray}
being $\s^{\pm}_l=(1/2)\left(\s^x_l \pm i \s^y_l\right)$ and
\begin{eqnarray}
K(l)= e^{i \pi \sum_{j=1}^{l-1} \s^{+}_j \s^{-}_j}= \prod_{j=1}^{l-1}
 (-\s^z_j).
 \label{vi}
\end{eqnarray}

These operators are true fermionic operators in the sense that
 $\left\{a^{\dagger}_l, a_m\right\}=\delta_{l,m}$ and 
$\left\{a^{\dagger}_l, a^{\dagger}_m\right\}=0=\left\{a_l, a_m\right\}$.
Furthermore, since
\begin{eqnarray}
\s^z_l=2 a^{\dagger}_la_l -1,
\label{sz}
\end{eqnarray}
the total third component of the states is, up to a constant factor,
 the number operator $N=\sum_{i=1}^{M}a^{\dagger}_ia_i$, and we can identify
 the states with $K$ spins flipped with those states that have $K$ particles
 of type $a$; furthermore, the total number of fermions is also a conserved
 quantity.
 Transforming from
 the $\s$  to the $a$ operators,
 the Hamiltonian can be written as
\begin{eqnarray}
H= &4& \sum_{l=1}^{M} { c_l(u,v)\left( N_l-{1\over 2}\right)
\left( N_{l+1}-{1\over 2}\right)} \nonumber \\
+&2& \sum_{l=1}^{M-1} { c_l(u,v) \left(a^{\dag}_l
a_{l+1}+a^{\dag}_{l+1} a_{l}\right)} - 2 e^{i \pi N} c_M(u,v) \left(a^{\dag}_M
a_{1}+a^{\dag}_{1} a_{M}\right),
\label{vii1}
\end{eqnarray}
being $N_l=a^{\dag}_l a_{l}$ and $c_l(u,v)=\left(u + (-1)^l v\right)$.
Due to the last term, the Hamiltonian reads different depending on the 
number of particles that has the state on which $H$  acts. Howewer, 
 considering separately the subspaces of even and odd total number of fermions,
 we may write $H$ in a simple universal way, that is
\begin{eqnarray}
H= u M+4 \sum_{l=1}^M {c_l(u,v)\left( N_l N_{l+1} - 
N_l\right)}+ 2 \sum_{l=1}^M { c_l(u,v) \left(a^{\dag}_l
a_{l+1}+a^{\dag}_{l+1} a_{l}\right)}
\label{vii}
\end{eqnarray}
where 
\begin{eqnarray}
a_{M+1} &=& - a_1, \,\,\,\,\, a_{M+1}^{\dagger} =- a_1^{\dagger}
 \,\,\,\,\text{ for states on which N is even} \nonumber  \\
a_{M+1} &=& a_1, \,\,\,\,\,\,\,\,\, a_{M+1}^{\dagger} = a_1^{\dagger}
 \,\,\,\,\,\,\,\, \text{ for states on which N is odd,} 
\label{bcf}
\end{eqnarray}
then, acting on states that have an even number of fermions the
 Hamiltonian (\ref{vii}) has antiperiodic boundary conditions, whereas 
 the boundary conditions are periodic when $H$ acts on states with an
 odd number of fermions. Trasforming to momentum space with
 the Fourier transformation
given by
\begin{eqnarray}
b_j & = & \frac{1}{\sqrt{M}}\sum_{l=1}^M e^{i k(j)l} a_l,\\
a_l & = & \frac{1}{\sqrt{M}}\sum_{j=1}^M e^{-i k(j)l} b_j,
\label{vft}
\end{eqnarray}
where 

 \[
  k(j)=\left\{ \begin{array}{ll}
  \frac{2 \pi}{M} \  j    & \mbox{if $m$ is odd},\\
  \frac{\pi}{M} \ (2 j -1) & \mbox{if $m$ is even}, \qquad j=1,\cdots,M
\label{ix}
  \end{array}
  \right. \]
and $m$ means the number of fermions of the subspace on which $H$
 acts; then the operators $b^{\dagger}_n$ and $b_n$ create or destroy,
 respectively,  a fermion of
 momentum $k(n)$ .
 With these transformations, the Hamiltonian (\ref{vii}) in the
momentum space is composed of  four pieces, plus the identity
times  $u M$,
\begin{eqnarray}
\label{xi}
H= H_1+H_2+H_3+H_4+ u M
\end{eqnarray}
where
\begin{eqnarray}
H_1&=&-4 u \sum_{n=1}^M{\varepsilon(n) b_n^{\dagger}b_n} \,,\\
H_2&=&-4 i v\sum_{n=1}^M {\sin(k(n)) b_{n+M/2} ^{\dagger}b_n} \,,\\
H_3&=&-4 {u\over M} \sum_{n_1,n_2,n_3=1}^M {e^{i(k(n_3)-k(n_1))} b_{n_1+n_2-n_3}^{\dag}
 b_{n_3}^{\dagger}b_{n_2} b_{n_1}} \,,\\
H_4&=&-4 {v\over M} \sum_{n_1,n_2,n_3=1}^M {e^{i(k(n_3)-k(n_1))}
b_{n_1+n_2-n_3+M/2}^\dag b_{n_3}^{\dagger}b_{n_2} b_{n_1}} \,,
\label{xii}
\end{eqnarray}
and
\begin{eqnarray}
\label{xiii}
 \epsilon (n)=1-\cos(k(n)).
\end{eqnarray}

The space of states will be generated now by a
 vacuum $|0>$ with zero
excitations, over which act the operators $b_l^\dag$ and $b_l$
that  create and destroy excitations that we call particles with
momentum $k(l)$; those states  are described by kets of the form
$|l_1,l_2,\cdots,l_m>$ that represents a state of $m$ particles
with momentums $k(l_1)$, $k(l_2),\cdots k(l_m)$. Obviously, the total
number of particles is conserved by every piece of the
Hamiltonian and then we can calculated the energy of the eigenstates with
a given number of particles.

The spectrum of energies in the one particle subspace can
 be obtained easily 
due to the fact that $H_1$ and $H_2$ 
are the only pieces that gives a  non null contribution.
Acting with $H$ on the states $ |l>$ and $|l+M/2>$, we have
\begin{eqnarray}\label{xiv}
H_1\, |l>& &= -4 u \,\epsilon (l) \,|l> \,, \\
H_2\,|l> & &= -4 i v \,\sin(k(l)) \,|l+\frac{M}{2}> \,,  \\
H_1\, |l+\frac{M}{2}>& & = -4 u\, \epsilon (l+\frac{M}{2}) \,|l+\frac{M}{2}> \,, \\
H_2 \,|l+\frac{M}{2}> & &= -4 i v\, \sin(k(l+\frac{M}{2}))\, |l> \,,
\end{eqnarray}
and $H_3\, |l>\,\,= \,H_4\, |l> \, \, =\,
H_3\, |l+M/2>\,\,= \,H_4\, |l+M/2> \, =0$.

Then we see that the two dimensional subspace 
 generated by the states $|l>$ and $|l+\frac{M}{2}>$
is invariant under the action of $H$ and consecuently the eigenvalues
 of $H_1 + H_2$ are those of the matrix,
\begin{eqnarray}
\left(\begin{array}{cc}
  -4 u \epsilon(l) & 4i v\sin(k(l)) \\
 -4i v\sin(k(l))  & -4 u \epsilon(l+\frac{M}{2})
\end{array} \right),
\label{xv}
\end{eqnarray}
and the total energies will be these eigenvalues plus $u\,M$, i.e.,
\begin{equation}\label{xvi}
E_l^{\pm} =\ u M-4 (u\pm \sqrt{(u\cos(k(l))^2+(v\sin(k(l))^2} ),
\quad l=1,\cdots,\frac{M}{2},
\end{equation}
where now $k(l)=(2 l \pi/M) $ as corresponds to the subspace
 with one particle. The corresponding eigenvectors are those of the
 matrix (\ref{xv})
\begin{eqnarray}
v_l^{\pm}={1 \over n_l^{\pm}} \,\,
\left( \begin{array}{c}
i (s_l\pm u \cos(k(l))) \\
\pm v\sin(k(l))
\end{array} \,\right),
\end{eqnarray}
where $s_l = [u^2 \cos^2(k(l))+ v^2 \sin^2(k(l))]^{1/2}$ and 
$ n_l^{\pm}= [2 s_l^2 \pm 2 s_l u \cos(k(l))]^{1/2}$.

\section{Two particles state case}

In this case, the pieces of the Hamiltonian
 $H_3$ and $H_4$ acting on the states are not zero
and it does more difficult to find the solution.
 The states are described by the kets of the form
$|l_1,l_2>$ with, $0<l_1< l_2\leq M$.
 We can observe that they are eigenstates of
 $H_1$ and that the action of  $H_3$ on them
 gives new states
$|l'_1,l'_2>$, in both cases they fulfill the condition
 $mod(l_1+l_2,M)= mod(l'_1+l'_2,M)$. In the
same form, the parts $H_2$ and $H_4$ mix the  states $|l_1,l_2>$
 with states $|l'_1,l'_2>$ being
$mod(l_1+l_2,M)= mod(l'_1+l'_2+ M/2,M)$. 
In other words, $H_1$ and $H_3$ conserve the total momentum whereas
 $H_2$ and $H_4$ change the total momentum by an amount of $\pi$.
 Then, the subspace generated by states
of the form
\begin{eqnarray}
|\Psi(m)> &=& \sum_{l={\rm All possibles}} g(l,m-l)|l,m-l> ,
\label {xviii1} \\
|\Psi(m+M/2)> &=& \sum_{l={\rm All possibles}} h(l,m-l+M/2)|l,m-l+M/2> ,
\label{xviii}
\end{eqnarray}
is invariant by the Hamiltonian, and we can reduce to this  subspace
 the problem of finding the eigenvalues and eigenvectors of $H$. Note that in 
 equations (\ref{xviii1}-\ref{xviii}) the functions $g(l_1,l_2)$ and 
 $h(l_1,l_2)$ are antisymmetric in their arguments, that is $g(l_1,l_2)
= -g(l_2,l_1)$ and $h(l_1,l_2)
= -h(l_2,l_1)$, in the same way  as they are the fermionic states
 $|l_1,l_2> = -|l_2,l_1>$. 

\subsection{Case $u=0$}

Let us begin with the case $u=0$. Then the total Hamiltonian
has only the alternating term so the equations are simpler and
 we can see beter the method. Working in units of $v$ (or 
  $v=1$) and taking
\begin{eqnarray}
\Xi(m)=i\psi(m)+\psi(m+M/2),
\label{esu0}
\end{eqnarray}
and demanding that $(H_2+H_4)\,\Xi(m) = E\,\Xi(m)$ we obtain that
\begin{eqnarray}
E\left[g(l,m-l)-g(m-l,l)\right]=\qquad \qquad \qquad \qquad \qquad
\qquad \qquad \qquad \qquad \qquad \qquad 
\nonumber  \\
8\left\{ 2B(m) \sin\left[k(l-m/2)\right]+
{1\over 2}\sin\left[k(l)\right]
\left( h(l-M/2,m-l)
-h(m-l,l+M/2)\right)+\right.\nonumber \\
+\left.\sin\left[k(m-l)\right]\left(h(l,m-l+M/2)-h(m-l-M/2,l)\right)\right\},
\label{l1}
\end{eqnarray}
and
\begin{eqnarray}
E\left[h(l,m-l+M/2)-h(m-l+M/2,l)\right]=
8\left\{ 2A(m) \cos\left[k(l-m/2)\right]+\right.
\qquad \qquad
\nonumber  \\
\left.
{1\over 2}\sin\left[k(l)\right]
\left( g(m-l+M/2,l-M/2)
-g(l-M/2,m-l+M/2)\right)+\right.\nonumber \\
+\left.\sin\left[k(m-l)\right]\left(g(l,m-l)-h(m-l,l)\right)\right\},
\label{l2}
\end{eqnarray}
where $A(m)$ and $B(m)$ are defined as
\begin{eqnarray}
A(m)&=&{1\over M}\sum_l {\sin\left[k(l-m/2)\right] \, g(l,m-l)},
\label{AB1}\\
B(m)&=&{1\over M}\sum_l {\cos\left[k(l-m/2)\right] \, h(l,m-l+M/2)}.
\label{AB2}
\end{eqnarray}

These equations (\ref{l1}) and (\ref{l2}) can be solved
 for $g$ and $h$. After a rather tedious calculus we have obtained
 the solutions for $g(l_1,l_2)$, $h(l_1,l_2)$ and the eigenvalues
 $E$ in the form
\begin{eqnarray}
& &h(l_1 ,l_2) =  {8 E A(l_1-l_2) \cos \left[ k((l_1-l_2+M)/2)\right] 
\over D_1(E,l_1,l_2)}+\nonumber \\
& &{64 B(l_1-l_2) \sin\left[ k((l_1+l_2+M)/2)\right]
\sin\left[ k((l_1-l_2+M)/2)\right]\cos\left[ k((l_1-l_2+M)/2)\right]
\over D_2(E,l_1,l_2)}, 
\nonumber\\
& &g(l_1 , l_2)= {8 E B(l_1-l_2) \sin \left[ k((l_1-l_2)/2)\right] 
\over D_1(E,l_1,l_2)}-\nonumber \\
& &\qquad\qquad{64 A(l_1-l_2) \cos\left[ k((l_1+l_2)/2)\right]
\sin\left[ k((l_1-l_2)/2)\right]\cos\left[ k((l_1-l_2)/2)\right]
\over D_2(E,l_1,l_2)},
\label{solg}
\end{eqnarray}
where the functions $D_i(E,l_1,l_2)$ are defined as
\begin{eqnarray}
D_i(E,l_1,l_2)&=& E^2 - 16 \left[ \sin(k(l_1))
-(-1)^i\sin(k(l_2))\right]^2, \,\,\,{\text{(i=1,2)}},
\label{D1}
\end{eqnarray}
and the eigenvalues $E$ depend only on the total momentum
 of the state $m=l_1+l_2$, and for each value of $m$ satisfy
 the consistency equation
\begin{eqnarray}
\left({8 E\over M}\right)^2 \, 
\left( \sum_l {\cos^2\left[k(l-m/2)\right] \over D_1(E,l,m-l)}
\right) \, \left( \sum_l {\sin^2\left[k(l-m/2)\right] \over D_2(E,l,m-l)}
\right)
=1.
\label{Evalue}
\end{eqnarray}

Notice that (\ref{Evalue}) is
 cuadratic in the energy $E$
 and hence it  gives symmetric eigenvalues
 $\pm E$ as corresponds to the charge conjugation
 symmetry of the Hamiltonian.
Note that this consistency equation (\ref{Evalue}) is of the type of 
 the eigenvalue equation for the Richardson's solution for the BCS model
\cite{rich}, so the study of the completeness of the solution is analogous
 to that of this model.
\begin{figure}[h!]
 \includegraphics[bb=40 0 467 240]{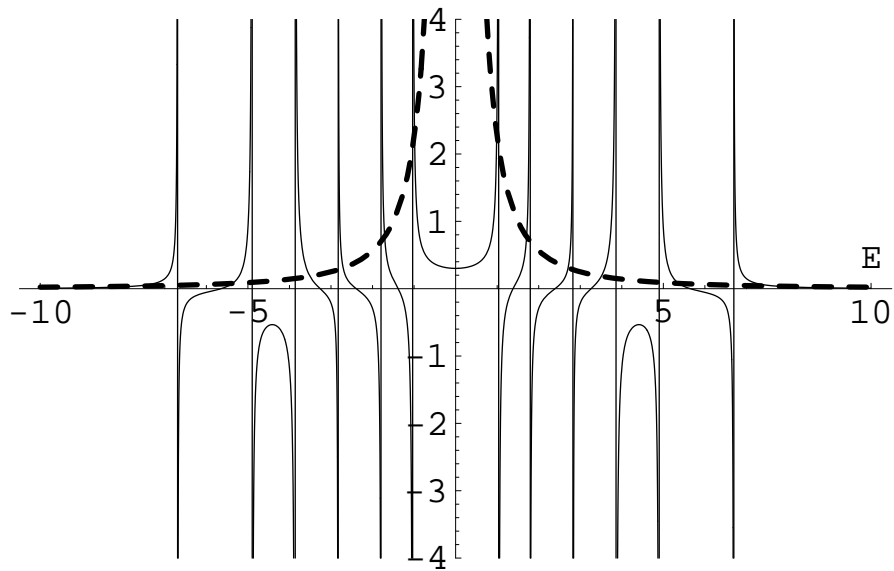}
\caption{\small Graphical solution of the equation (\ref{Evalue})
for $m=3$ and $M=12$.
Dashed line corresponds to the $G(E)=(8 E/M)^{-2}$ term and the continuus
line is  F(E).  } 
\label{fig-ejemplo}
\end{figure}

 As an example, in  Fig. (\ref{fig-ejemplo}) we plot
the graphical solution of Eq. (\ref{Evalue}) for the case $M=12$ and $m=3$; there we plot the curves $G(E)=(8 E/M)^{-2}$ and 
$F(E)=\left( \sum_l {\cos^2\left[k(l-m/2)\right] \over D_1(E,l,m-l)}
\right) \, \left( \sum_l {\sin^2\left[k(l-m/2)\right] \over D_2(E,l,m-l)}
\right)$. Here it is evident that the number of solutions is determined
 by the discontinuities of the function $F(E)$ along the positive axis.

\subsection{General case}

Now we return to the general case when $u$ and $v$ are $\ne 0$. 
The action of the different parts of the Hamiltonian on the states  
 (\ref{xviii1} - \ref {xviii}) is given by,
\begin{eqnarray}
\label{xix}
&H_1|\psi(m&)> = - 4 u \sum_l^{}(\varepsilon(l)+
\varepsilon(m-l))\,g(l,m-l)\,|l, m-l>, \\
&H_1|\psi(m&+M/2)>= - 4 u \sum_l^{}(\varepsilon(l)+
\varepsilon(m-l+M/2))\nonumber \\
& &\ \ \ \ \ \ \ \ \ \ \ h(l,m-l+M/2)\,|l, m-l+M/2>, \\  
&H_2|\psi(m&)> = -  4 i v \sum_l^{}(-\sin(k(l))\,g(l-M/2,m-l+M/2)\,
\nonumber \\
& &\ \ \ \ \ \ \ \ \ \ \ \ +\sin(k(m-l))\,g(l,m-l))\,|l, m-l+M/2>, \\
&H_2|\psi(m&+M/2)> =   4 i v \sum_l^{}(\sin(k(l))\,h(l-M/2,m-l)\,
\nonumber \\
 &  &\ \ \ \ \ \ \ \ \ \ \  +\sin(k(m-l))\,h(l,m-l+M/2))\,|l, m-l>, \\
&H_3|\psi(m&)> =  16 u A(m) \sum_l^{} \sin(k(l)-s(m)/2)\,|l, m-l>, \\
&H_3|\psi(m&+M/2)>= 16 u B(m) \sum_l^{}\cos(k(l)-s(m)/2)\,|l, m-l+M/2)>,\\
&H_4|\psi(m&)> = 16 i v  A(m) \sum_l^{}\cos(k(l)-s(m)/2)\,|l, m-l+M/2)>, \\
&H_4|\psi(m&+M/2)> =  16 i v B(m) \sum_l^{} \sin(k(l)-s(m)/2)\,|l, m-l>,
\end{eqnarray}
being now 
\begin{eqnarray}
\label{xx}  
A(m) &=& \frac{1}{M}\sum_l \sin(k(l)-s(m)/2) g(l,m-l),\label{xxa}  \\
B(m) &=& \frac{1}{M}\sum_l \cos(k(l)-s(m)/2) h(l,m-l+M/2),\label{xxb} 
\end{eqnarray}
and
\begin{equation}
  s(m)=\frac{2\pi}{M}(m-1).
\label{xxi} 
\end{equation}

 The sums are understood to all different values of the index
  and the integer labels
 of the states  are module $M$.

The energies will be obtained from the compatibility of the eigenvalue
 equation
 \begin{equation}
   (H_1+H_2+H_3+H_4)(\Psi(m)+\Psi(m+M/2))=E \,(\Psi(m)+\Psi(m+M/2)).
\label{xxii}
 \end{equation}

Equating the coefficients of every state in the first
 member and in the second member, we can observe that 
the relations between them mix only the 
$g(l,m-l), g(l-M/2,m-l+M/2)$, $ h(l-M/2)$ and $h(l-M/2,m-l)$ terms,
besides of the functions $A(m)$ and $B(m)$ that depends of
 all coefficients. This relations are expressed as the
 system of linear equations,
 \begin{equation}
   (E\mathbf{I}-\mathbf{M})\left( \begin{array}{c} g(l,m-l)\\ g(l-\frac{M}{2},m+\frac{M}{2}-l)\\h(l,m+\frac{M}{2}-l)\\h(l+\frac{M}{2},m-l)
\end{array} \right)=
16 \left( \begin{array}{c}\sin(k(l)-\frac{s(m)}{2})(u A(m)+i v B(m)) \\
-\sin(k(l)-\frac{s(m)}{2})(u A(m)+i v B(m)\\
\cos(k(l)-\frac{s(m)}{2})(-iv A(m)+u B(m))\\
-\cos(k(l)-\frac{s(m)}{2})(-iv A(m)+u B(m))
\end{array} \right)
\label{xxiii}
 \end{equation}
where $\mathbf{M}$ is the matrix
 \begin{equation}
\mathbf{M}=
 \left( \begin{array}{cccc}
d_1(l,m) & 0 & c_1(l,m) & c_2(l,m) \\
0  & d_1(l-\frac{M}{2},m) & -c_2(l,m)&-c_1(l,m)  \\
-c_1(l,m) & c_2(l,m) & d_2(l,m) & 0  \\
-c_2(l,m) & c_1(l,m) & 0 & d_2(l-\frac{M}{2},m)
\end{array} \right)
\label{xxiv}
 \end{equation}
with 
\begin{eqnarray}
\label{xxv}  
&c_1(l,m)&\,=\, 4 i v \sin(k(m-l)) \ ,\\
&c_2(l,m)&\,=\, 4 i v \sin(k(l))  \ ,\\
&d_1(l,m)&\,=\, -4 u (\varepsilon(l)+\varepsilon(m-l)) \ ,\\
&d_2(l,m)&\,=\, -4 u (\varepsilon(l)+\varepsilon(m-l+\frac{M}{2})). \
\end{eqnarray}

From the equations (\ref{xxiii}), we can get the solutions of $g(l,m-l)$ and 
$h(l,l-m+M/2)$ by making use of the inverse matrix of $(E\mathbf{I}-
\mathbf{M})$. The matrix elements of that matrix are function of
 the $l$, $m$, 
and the eigenvalues $E$ of 
the Hamiltonian $H=H_1+H_2+H_3+H_4$. 
If we call $s_{i,j}$ to the matrix
 elements of $(E\mathbf{I}-\mathbf{M})^{-1}$ we obtain,
\begin{eqnarray}
\label{xxvi}  
&g(l,m-l&)=
16( \cos(k(l)-\frac{s(m)}{2})(s_{1,3}-s_{1,4})(-i v A(m)+u B(m)) \nonumber \\
 & & + \sin(k(l)-\frac{s(m)}{2})(s_{1,1}-s_{1,2}))(u A(m)+i v B(m)))\ ,\label{xxvia} \\
&h(l,m-l&+\frac{M}{2})=
16( \cos(k(l)-\frac{s(m)}{2})(s_{3,3}-s_{3,4})(-i v A(m)+u B(m)\nonumber \\
& & + \sin(k(l)-\frac{s(m)}{2})(s_{3,1}-s_{3,2}))(u A(m)+i v B(m))) \ ,
 \label{xxvib}
\end{eqnarray}
where the $s$ functions are
\begin{eqnarray}
s_{1 1}& = &\frac{1}{D}\,{c_1(l,m)}^2
   \left( E -{d_2}(l,m) \right)  +   \nonumber\\
& &\frac{1}{D} \left( {{c_2}(l,m)}^2
    + \left( E -{d_1}(l -
          \frac{{M}}{2},m) \right) \,
      \left( E - {d_2}(l,m) \right)  \right) \,
   \left( E -{d_2}(l -
       \frac{{M}}{2},m) \right)  \nonumber  \\
s_{1  2}& = &\frac{1}{D}c_1(l,m)
\,{c_2}(l,m)\,
  \left( 2\,E - {d_2}(l,m) -
    {d_2}(l - \frac{{M}}{2},m)
    \right)  \nonumber \\
s_{1 3}& = & \frac{1}{D}\,{c_1}(l,m)
  \left( {{c_1}(l,m)}^2 - 
    {{c_2}(l,m)}^2
+\left( E - {d_1}(l -
         \frac{{M}}{2},m) \right) \,
     \left( E - {d_2}(l - 
         \frac{{M}}{2},m) \right)  \right)  \nonumber\\
s_{1 4}& = & \frac{1}{D}{c_2}(l,m)
\,
  \left( -{{c_1}(l,m)}^2 +
    {{c_2}(l,m)}^2 +
    \left( E -{d_1}(l -
         \frac{{M}}{2},m) \right) \,
     \left( E - d_2(l,m) \right)  \right)  \nonumber\\
s_{2 2}& = & \frac{1}{D}
\left( E - {d_2}(l,m) \right) \,
   \left( {{c_2}(l,m)}^2 + 
     \left( E -{d_1}(l,m) \right) \,
      \left( E - {d_2}(l - 
          \frac{{M}}{2},m) \right)  \right)  +\nonumber \\
 &  &  \frac{1}{D}{{c_1}(l,m)}^2\,
   \left( E - {d_2}(l -
       \frac{{M}}{2},m) \right)  \nonumber \\
s_{2 3}& = & \frac{1}{D}
{c_2}(l,m)\,
  \left( {{c1}(l,m)}^2 -
    {{c_2}(l,m)}^2 - 
    \left( E - {d_1}(l,m) \right) \,
     \left( E - {d_2}(l - 
         \frac{{M}}{2},m) \right)  \right)) \nonumber \\
s_{2 4}& = & \frac{1}{D}
{c_1}(l,m)\,
  \left( -{{c_1}(l,m)}^2 + 
    {{c_2}(l,m)}^2 -
    \left( E - {d_1}(l,m) \right) \,
     \left( E - {d_2}(l,m) \right)  \right) \nonumber  \\
s_{3 3}& = & \frac{1}{D}
{{c_1}(l,m)}^2\,
   \left( E - {d_1}(l,m) \right)  + 
 \frac{1}{D}  \left( E - {d_1}(l -
       \frac{{M}}{2},m) \right) \, \nonumber  \\
   &  &\left( {{c_2}(l,m)}^2 +
      \left( E - {d_1}(l,m) \right) \,
      \left( E - {d_2}(l -
          \frac{{M}}{2},m) \right)  \right)\nonumber  \\
s_{3 4}& = & \frac{1}{D}
{c_1}(l,m)\,{c_2}(l,m)\,
  \left( -2\,E + {d_1}(l,m) +
    {d_1}(l - \frac{{M}}{2},m)
    \right)  \nonumber  \\
s_{4 4}& = & \frac{1}{D}
{{c_1}(l,m)}^2\,
   \left( E - {d_1}(l -
       \frac{{M}}{2},m) \right)   \nonumber  \\
&  &\hbox{    }+  \frac{1}{D} \left( E - {d_1}(l,m) \right) \,
   \left( {{c_2}(l,m)}^2 +
     \left( E - {d_1}(l -
          \frac{{M}}{2},m) \right) \,
      \left( E - {d_2}(l,m) \right)  \right)
\end{eqnarray}
being $D$ the determinant of $ E {\bf I}-{\bf M}$, i.e.,
 the characteristic polynomial of ${\bf M}$
\begin{eqnarray}
D &= &
{{c_1}(l,m)}^2\,\left( {{c1}(l,m)}^2 - {{c_2}(l,m)}^2 + 
     \left( E - {d_1}(l,m) \right) \,\left( E - {d_2}(l,m) \right)  \right)  \nonumber \\
  & &    + {{c_2}(l,m)}^2\,\left( -{{c_1}(l,m)}^2 + 
     {{c_2}(l,m)}^2 + \left( E -
        {d_1}(l - \frac{{M}}{2},m) \right) \,
      \left( E - {d_2}(l,m) \right)  \right)  \nonumber \\
 & & +   \left( {{c_2}(l,m)}^2\,\left( E - {d_1}(l,m) \right)  + 
     \left( E - {d_1}(l - \frac{M}{2},m) \right) \, \right.  \nonumber \\
 & & \left.   \left( {{c_1}(l,m)}^2 + 
        \left( E - {d_1}(l,m) \right) \,\left( E - {d_2}(l,m) \right)
        \right)  \right)
 \,\left( E - {d_2}(l - \frac{{M}}{2},m) \right)
\end{eqnarray}

With the last expressions, we can substitute $g(l,m-l)$
 and $h(l,m-l+M/2)$ given by equations  (\ref{xxvi})- (\ref{xxvib})
 in the definitions for $A(m)$ and $B(m)$ given in the
 equations (\ref{xxa},\ref{xxb}). If we call
\begin{eqnarray}
\label{xxva}  
c_{1,1}(m,E)&=&\frac{16}{M}\sum_{l} \sin( k(l)-\frac{s(m)}{2}) \cos( K(l)-\frac{s(m)}{2}(s_{1,3} -s_{1,4}), \\
c_{1,2}(m,E)&=&\frac{16}{M}\sum_{l}  \sin( k(l)-\frac{s(m)}{2})^{2}(s_{1,1}-s_{1,2}),\\
c_{2,1}(m,E)&=&\frac{16}{M}\sum_{l}  \cos( k(l)-\frac{s(m)}{2})^{2}(s_{3,3}-s_{3,4}), \\
c_{2,2}(m,E)&=&\frac{16}{M}\sum_{l} \sin( k(l)-\frac{s(m)}{2}) \cos( K(l)-\frac{s(m)}{2})(s_{3,1} -s_{3,2}),
\end{eqnarray}
Then we obtain the equations
\begin{eqnarray}
\label{xxvii}
A(m)&=& c_{1,1}(m,E) (-i v A(m) + u B(m))+ c_{1,2}(m,E) (u A(m) +i v B(m)) \\
B(m)&=& c_{2,1}(m,E) (-i v A(m) + u B(m))+ c_{2,2}(m,E) (u A(m) +i v B(m)). 
\end{eqnarray}
The compatibility of that system impose the condition,
\begin{eqnarray}
\label{xxviii}
\left| \begin{array}{cc}
-i v c_{1,1}(m,E)+u c_{1,2}(m,E)-1&u c_{1,1}(m,E)+i v c_{1,2}(m,E)\\
-i v c_{2,1}(m,E)+u c_{2,2}(m,E) & u c_{2,1}(m,E)+i v c_{2,2}(m,E)-1
\end{array}\right| = 0,
\end{eqnarray}
this  equation plays the role of the Bethe ansatz equations 
and it can be solved by algebraic or numerical methods 
for any $1\,\leq m\,\leq \,M$ and an arbitrary number
of sites $M$. 

The determinant $D$ is the four degree 
characteristic polynomial of the matrix 
${\bf M}$ that correspond to the Hamiltonian 
piece $H_{12}=H_1+H_2$. It is factorized as the product
\begin{equation}
D(E)\,=\, \prod_{i=1}^{4} (E\,-\,E_i)
\end{equation}
where $E_i$ are the eigenvalues of $H_{12}$. In some cases it can 
happen that some  of this factors will be common to all 
numerators of the functions $g$ and $h$ involved in the
 same set of equations (\ref{xxiii}).  Then, the corresponding
 eigenvalue $E_i$ of $ H_{12}$ will be  also eigenvalue of the total
 Hamiltonian, i.e., solution of (\ref{xxviii}). Due to that, we
 must take care since that eigenvalue will not appear if we  do 
the simplification in the expressions for the $g$ and $h$ functions.
 
For small $M$ the solutions of the equation (\ref{xxviii})
 can be compared with the solutions obtained after diagonalizing 
the total Hamiltonian in both, the site space or the momentum space.
 For example, for $M=6$ we have  that the dimension of the state space 
 is  15 whereas the invariant subspace generate by the states with 
$m$ and $m+M/2$ has $5$ dimensions. The solutions can be obtained 
by algebraic methods in the two cases. 
Using the Mathematica application, for $m=1$,
 we obtain in the two cases the solutions

\begin{eqnarray}
&E=& \,0 \label{uno} \\
&E=& \,-8\,u \\
&E=&-8\,u + \frac{16}{(12)^{1/3}} \,\frac{(u^2 + v^2)}
    {f(u,v)} + 2\, \left( \frac {2}{3} \right)^{\frac {2}{3}}\, f(u,v)\\
&E=&
-8\,u - \frac{8\,\left( 1 + i \,{\sqrt{3}} \right) }{(12)^{1/3}} \,\frac{ \,\left( u^2 
+ v^2  \right) }
   {f(u,v)} -
  {\left( \frac{2}{3} \right) }^{\frac{2}{3}}\,
\left( 1 -i \,{\sqrt{3}} \right) \,
  f(u,v) \\
&E=&
-8\,u - \frac{8\,\left( 1 - i \,{\sqrt{3}} \right) }{(12)^{1/3}} \,\frac{ \,\left( u^2 
+ v^2  \right) }
   {f(u,v)} -
  {\left( \frac{2}{3} \right) }^{\frac{2}{3}}\,
\left( 1 +i \,{\sqrt{3}} \right) \,
  f(u,v), 
\label{xxx}
\end{eqnarray}
where $f(u,v)$ is defined as
\begin{eqnarray}
f(u,v)=\left( -9 u\,(u^2 - v^2) + {\sqrt{3}}\,
        \sqrt{-5\,u^6 - 150\,u^4\,v^2 - 69\,u^2\,v^4 - 32\,v^6}
 \right)^{\frac{1}{3}}
\end{eqnarray}

The solution $E\,=\,- 8u $ is also eigenvalue of $H_{12}$.
Numerical evaluations show that these solutions are real as correspond to an selfadjoint Hamiltonian.

It is interesting to get the values of the  energies for the special
 cases $v=0$ and $u=0$. In the first case, we obtain obviously the
 spectrum of the $XXX$-Heisenberg model and the equations (\ref{xxiii})
 become very simple since $H_1$ is diagonal and $H_2$ is null. In the second
 case, $u=0$, the equations (\ref{xxiii}) are simplified considerably
 too, since now the functions $d$ are null. In both cases the solutions are,
\begin{eqnarray}
E_{v=0}&=&\{0,\,-8\,u,\, -4 u,\, 2 
\left(-5+\sqrt{5}\right)u,\,-2 \left(5+\sqrt{5}\right) u\}, \qquad\\
E_{u=0}&=&\{0,\,0,\,0,\,4\sqrt{2}v,\,-4\sqrt{2} v\}, \qquad
\end{eqnarray}
that have been obtained from (\ref{uno} - \ref{xxx}) by making use
 of the respective 
substitutions for $u$ and $v$.

{\bf Acknowledgments}

We thank to M. F. Ra\~nada for comments.
This work has been supported the
CICYT  grants  BFM2000-1057
and FPA2000-1252.



\begin{thebibliography}{99}

\bibitem {spchain} {J. W. Bray, H. R. Hart, Jr, L. V. Interrante,
 I. S. Jacobs, J. S. Kasper, G. D. Watkius, S. H. Wee and J. C. Bonner,
Phys. Rev. Lett. {\bf 35}, 744 (1975). \\
     M. Hase, I. Terasaki and K. Uchinokura, 
Phys. Rev. Lett. {\bf 70}, 3651 (1993).\\
M. Hase, I. Teraski, K. Uchinokura, M. Tokunaga, 
N. Miura and H. Obara, Phys. Rev. {\bf B48}, 9161 (1993).\\
J. P. Boucher and L.P. Regnault, J. Physique {\bf I6}, 1939 (1996).\\
D. C. Johnston, J. W. Johnson, D. P. Gorshorn and A. J. Jacobson, 
Phys. Rev. {\bf B35}, 219 (1987).\\
H. P. Nightingale and H. W. Bl\" ote, Phys. Rev. {\bf 33}, 659 (1986).\\
S. R. White and D. A. Huse, Phys. Rev. {\bf B48}, 3844 (1993).}

\bibitem {htc} {E. Dagotto, Int. J. Mod. Phys. {\bf C 2}, 659 (1991) \\
E. Manousakis, Rev. Mod. Phys. {\bf 63}, 1 (1991).}

\bibitem
{r3}{H. Bethe, Z. Physik {\bf 71}, 205 (1996).

It can be found too in many textbooks, for example, 
E. Fradkin, {\it Field Theories of Condensed Matter Systems}
 (Addison-Wesley Publishing Co. New York, U.S.A. 1991).}

\bibitem
{r4}{E. K. Sklyianin, L.A. Takhtajan and L. D. Fadeev,
 Ther. i. Mate. Fisika {\bf 40}, 194 (1979). 

For a review, V. E. Korepin, N. M. Bogoliubov 
and A.C.Izergin, {\it Quantum Inverse Scattering 
Method and Correlations Functions} (Cambridge University
Press, Cambridge, U.K., 1993) and Z. N. Ha 
{\it Quantum Many Body Systems in One Dimension} 
(World Scientific Publishing, Singapore (1996).}

\bibitem
{r5}{M. C. Cross and D. S. Fisher, Phys. Rev. B {\bf 19}, 402 (1979).

T. Nakano and H. Fukuyama, J. Phys. Soc. Jpn. {\bf 49}, 1679 (1980).}

\bibitem
{r6}{H. Frahm and J. Schliemann, Phys. Rev. B {\bf56}, 5359 (1997).}

\bibitem
{gog}{A. O. Gogolin, A. A. Nersesyam and A.M. Tsvelik, {\it
Bosonization and Strongly Correlated Systems} (Cambridge University
Press, Cambridge, U.K. 1998).}

\bibitem{rich} {R. W. Richardson, Phys. Lett. {\bf 3}, 277 (1963) \\
R. W. Richardson and N. Sherman, Nucl. Phys. {\bf B 52}, 221 (1964).}
 \end{thebibliography}
\end{document}